\documentclass[fleqn,twoside]{article}
\usepackage{amsmath}
\usepackage{espcrc2}

\newcommand{\ed}{\end{document}}
\newcommand{\be}{\begin{equation}}
\newcommand{\ee}{\end{equation}}
\newcommand{\ea}{\end{eqnarray}}
\newcommand{\ba}{\begin{eqnarray}}
\newcommand{\nn}{\nonumber\\}
\newcommand{\dslash}{\partial\!\!\!/}
\newcommand{\aslash}{A\!\!\!/}

\newcommand{\fslash}{f\!\!\!/}
\newcommand{\Dslash}{D\!\!\!\!/}

\title{Duality and Topological Mass Generation in Diverse Dimensions}

\author{C. Wotzasek\address{Instituto de F\'\i sica\\ 
Universidade Federal do Rio de Janeiro\\ 
21945 Rio de Janeiro}%
        \thanks{This work is supported in part by funds provided by the CNPq, CAPES and FAPERJ}}
       
\begin{document}

\begin{abstract}
We shall discuss issues of duality and topological mass generation in diverse dimensions. Particular emphasis will be given to the mass generation mechanism from interference between self and anti self-dual components, as disclosed by the soldering formalism. This is a gauge embedding procedure derived from an old algorithm of second-class constraint conversion used by the author to approach anomalous gauge theories. The problem of classification of the electromagnetic duality groups, both massless and massive, that is closely related will be discussed. Particular attention will be paid to a new approach to duality based on the soldering embedding to tackle the problem of mass generation by topological mechanisms in arbitrary dimensions including the couplings to dynamical matter, nonlinear cases and nonabelian symmetries.
\vspace{1pc}
\end{abstract}

\maketitle

\section{Preliminaries.}
Combining different species of objects
to yield a new composite structure is a very recurrent concept in Physics and Mathematics. 
The fusion process can be established as long as the constituents display either opposite or complementary aspects of some definite property.  This
simple observation contains the genesis of the technique, which
we call the Soldering Formalism. 
Over the last few years we have systematically developed the soldering
formalism and extensively applied to a wide variety of problems,
including higher dimensional models. The contribution of our group to this program is listed below \cite{ADW,ABW,BW,CW27,AW26,CW23,AINW,BW17,WI14,IW13,BINRW,IW11,AINRW,IW8,AsDW7,MNRW6,MNRW3,AsDW3}.

The soldering mechanism is a new technique developed to work with distinct
manifestations of self dual aspects of some symmetry, for instance,
chirality, helicity and electromagnetic self-duality, which can be soldered.
It provides a clean algorithm, inherited from our constraint conversion program\cite{constraints}, for fusing the opposite nature of these
symmetries by taking into account interference effects.
Therefore, two models carrying the representation of such symmetries being
otherwise completely independent, can be fused by this technique,
irrespective of dimensional considerations.
It can be applied to the quantum mechanical harmonic
oscillator\cite{chiraloscil} or used to investigate the cases of
electromagnetic dualities in distinct dimension, including the new and interesting instance of non commutative manifolds.
The list of applications is indeed quite extensive. 

After discussing the general aspects of the method we shall focus
on a few applications. Although our initial study in soldering was to investigate the possibility of fusing different chiralities of two independent systems displaying truncated diffeomorphism into a 2D gravity model, we shall concentrate on two different lines of developments.  One is to explore the intimate connection between soldering and duality to study the phenomenon of topological mass generation. The other is to study the possibility of simultaneously considering the soldering and bosonization, in different
dimensions. 

An interpretation of the soldering process has also been offered that discloses the whole process as a canonical transformation, however in the Lagrangian side\cite{BG}. We coined it dual projection. This new interpretation has provided us with some practical applications and used to study some controversial issues to date, such as the way to properly couple chiral matter to gauge and gravitational backgrounds\cite{couplings} or to establish to equivalence between different formulations of chiral modes besides providing new interpretation to their field constituintes. 
It also offered the opportunity to treat the electromagnetic duality groups under this technique and to disclose their group structure dimensional dependence, both for massless and massive fields and to propose a new physical interpretation for the phenomenon\cite{CW23}.  It also opens up the possibility to deal with electromagnetic duality for arbitrary p-form theory in different dimensions.  In a different line of investigation, the formalism was also used to study the duality equivalence of related models, including nonlinear effects, in distinct dimensions and couplings to different matters\cite{WI14,BINRW,IW11,AINRW,IW8,MNRW6,MNRW3}.

\section{Soldering and Canonical Transformations.}
The basic idea is to raise a global Noether
symmetry of the self and anti-self dual constituents into a
local one, but for an effective composite system, consisting of the
dual components and an interference term that defines the
soldered action.  Here we adopt an iterative Noether procedure
to lift the global symmetries.  Assume the symmetries
are being described by the local actions
\be
\label{ii05}
S^{(0)}_{\pm}(\varphi_{\pm}^\eta)=\int \, L^{(0)}_{\pm}(\varphi_{\pm}^\eta)\, dt
\ee
invariant under a global multi-parametric transformation $\delta \varphi_{\pm}^\eta = \alpha^\eta$ 
where the suffix (0) indicates the
iterative nature of the analysis.
Here $\eta$ represents the
tensorial character of the basic fields and, for simplicity, will be dropped from now on.
In general, under  local
transformations these actions will not remain invariant,
and Noether counter terms
become necessary in order to reestablish the invariance.
These counter terms contain, apart from the original fields, appropriate
contributions of auxiliary fields $W^{(N)}$.
Thus, after N iterations we obtain, 
\be
\label{ii20}
S_{\pm}^{(0)}\rightarrow S_{\pm}^{(N)}=
S_{\pm}^{(N-1)}- W^{(N)} J_{\pm}^{(N)}
\ee
where $J_{\pm}^{(N)}$ are the Noether currents. For the individual self and anti-self dual
systems this iterative procedure will not produce gauge invariant
systems.  However, for the composite system this procedure will, eventually, lead 
to an effective action of the form,
\ba
\label{ii21}
\!\!\!S(\varphi_\pm,W)^{(N)}\!\!\!&=&\!\!S_{+}(\varphi_{+})^{(N-1)}\!\!+\!\!S_{-}
(\varphi_{-})^{(N-1)}\nn&+&\!\! W^{(N-1)}\!\!\left(\!J_{+}^{(N-1)}+J_{-}^{(N-1)}\!\right)\!\!
\ea
which turns out to be 
invariant under the original transformations. The auxiliary fields $W^{(N)}$ are then
eliminated in favor of the other variables.
The final effective action then becomes a function of the original variables
only. In this form the effective action is no longer a function of the
individual fields but of some composite as $\phi=\varphi_+-\varphi_-$ if the fields belong to the algebra of some group or $\phi =\varphi_+ \cdot \varphi_-$ if they are group variables,
\be
\label{ii23}
S(\varphi_\pm,W)^{(N)}{\mid}_{W=f(\varphi_\pm)}
\to S_{eff}(\phi)
\ee
so that, $\delta S_{eff} = 0$.
This is
our cherished effective action which is a result of the soldering or the
fusion of the original actions.

\section{Dynamical Mass Generation and Interference.} 
Here we discuss the applications of the 
soldering technique to the problems of quantum field theory where the full 
power of this approach is manifested. Look for original papers for quantum mechanics\cite{chiraloscil} and applications\cite{applications}.
We discuss first two-dimensional field theory
where the use of the bosonization technique provides us with exact results.  
The extension to the non Abelian example
and a new look at the Polyakov-Weigmann identity\cite{ABW}, as well as the result of different regularization schemes, known to exist in the chiral Schwinger models
leading to distinct categorization of second class constraints, has been proposed\cite{AINW}. 

Three dimensional models are considered in the sequel.
Although bosonization is not exact in these dimensions, yet some expressions are known
in particular limiting conditions. Our interest is basically confined to the
long wavelength limit where local expressions are available. Under these conditions 
the massive Thirring model or quantum electrodynamics bosonize to self dual models.
Soldering of such models is carried out to get new models. 
The soldering of models with different coupling parameters
is given.  Three dimensional gravity has been examined in the literature where the
soldering of linearized Hilbert-Einstein actions with Chern-Simons terms is studied\cite{IW13}.
Some attention has also been devoted to a study of W gravity models\cite{AsDW7}. The $W_2$ and $W_3$ cases, which yields
exact soldering, were discussed in details.
Finally the case of higher conformal spins $W_n (n>2)$, which is developed within some perturbative scheme since soldering is not exact in this instance, was considered.
We will not give details of all these examples in this review.

\subsection{Soldering and 2D Bosonization.}
Bosonization is a powerful technique that  maps a
fermionic theory into its bosonic counterpart. 
It was fully explored in the context of two dimensions\cite{AAR} and later
extended to higher dimensions\cite{M,C,RB,RB1}.
Its importance lies in
the fact that it includes quantum effects already at the classical level. 
Consequently, different aspects and manifestations of quantum phenomena
may be investigated directly, that would otherwise be highly nontrivial
in the fermionic language.

Consider an
explicit one loop calculation following Schwinger's point splitting method\cite{J}
which is known to
yield\cite{RB2},
\begin{eqnarray}
\label{30s}
&&W_+[\omega] = -i \log \det (i\dslash+e\aslash_\pm)\\
&=& \left\langle\partial_+
\omega\partial_-\omega +2 \, e\,A_\pm\partial_\mp\omega + a_\pm\, 
e^2\, A_+ A_-\right\rangle\nonumber
\ea
where $\omega = \varphi,\rho$ and $\langle\cdots\rangle$ means spacetime integration. Here light cone metric is used and the regularization ambiguity is manifested through $a_\pm$. In the Hamiltonian approach their values define a
second-class constrained systems with two and four constraints.

Here different scalar fields $\varphi$ and $\rho$ are used to emphasize that the fermionic chiral components are uncorrelated.
It is the soldering that abstracts a meaningful combination, this being essentially the
simultaneous gauging of a global symmetry of the individual chiral components.
Consider, therefore,
the gauging of the following global symmetry, $\delta \varphi = \delta\rho=\alpha$,
$\delta A_{\pm}= 0$. The variations in the effective actions  (\ref{30s}) are found to be,
\begin{eqnarray}
\label{50s}
\delta W_\pm[\eta] = \left\langle \partial_\mp\alpha \;J_\pm(\eta)\right\rangle
\end{eqnarray}
\noindent  where the currents are defined as,
$J_\pm(\eta)={1\over{2\pi}}(\partial_\pm\eta +\, e\,A_\pm)$; $\eta= \varphi , \rho$.  Next we introduce the 
soldering field $B_\pm$ so that,
\begin{equation}
\label{70s}
W_\pm^{(1)}[\eta] = W_\pm[\eta] - \left\langle B_\mp\, J_\pm(\eta)\right\rangle
\end{equation}
\noindent Then it is possible to define a modified action,
\begin{equation}
\label{80s}
W[\varphi,\rho]= W_+^{(1)}[\varphi] + W_-^{(1)}[\rho]
 + \left\langle B_+ \,B_-\right\rangle
\end{equation}
\noindent which is invariant under an extended set of transformations that 
includes the matter fields together with, $\delta B_{\pm}= \partial_{\pm}\alpha$. Observe that the soldering field transforms as a vector potential.
Since it is an auxiliary field, it can be eliminated from (\ref{80s}). This will naturally solder the
otherwise independent chiral components. The relevant solution is found to be,
$B_\pm= 2\, J_\pm$. Inserting this solution in (\ref{80s}), we obtain,
\begin{eqnarray}
\label{110s}
W[\Phi] = \langle \left\{ \left( \partial_+\Phi \partial_-\Phi + 2e A_+\partial_-\Phi
- 2e A_- \partial_+\Phi \right)\right. &&\nn
 +\left.(a_+ + a_- -2)\,e^2\,A_+\,A_-\right\}\rangle &&
\end{eqnarray}
\noindent where, $\Phi=\varphi - \rho$. The action is no longer expressed in terms of the
different scalars $\varphi$ and $\rho$, but only on their gauge invariant combination.

Let us digress on the significance of the findings. At the classical fermionic
version, the chiral Lagrangians are completely independent. Bosonizing them
includes quantum effects, but still there is no correlation. The soldering
mechanism exploits the symmetries of the independent actions to
precisely combine them to yield a single action.
Note that the soldering works with the bosonized expressions. Thus the soldered
action obtained in this fashion corresponds to a new quantum theory.

We now show that different choices for the parameters $a_\pm$ 
lead to well known models.  To do this consider the variation of
(\ref{110s}) under the conventional gauge transformations.
Since the expression
in parenthesis is gauge invariant, a gauge invariant
structure for $W$ is obtained provided, $a_+ + a_- -2=0$.
By functionally integrating
out the $\Phi$ field from (\ref{110s}), we obtain,
\begin{equation}
\label{140s}
W[A_\pm]=  { e^2\over 4\pi}\left\langle A_+ 
{\partial_-\over \partial_+}A_+ 
\!+\! A_- {\partial_+\over \partial_-}A_- \!-\! 2 A_+ A_-\right\rangle
\end{equation} 
which is the familiar structure for the gauge invariant action expressed in
terms of the potentials. The opposite chiralities of the 
two independent fermionic theories have been soldered to yield a gauge 
invariant action.

Interesting observations are possible concerning the regularization
ambiguity parameters $a_\pm$. A single equation
cannot fix both the parameters but Bose symmetry imposes a crucial restriction\cite{BB}.
This symmetry plays
an essential part that complements gauge invariance. In the present case,
this symmetry corresponds to the left-right (or + --) symmetry in (\ref{30s}),
thereby requiring $a_+ = a_- = 1$. This has important consequences if a Maxwell term
was included from the beginning to impart dynamics. Then the soldering takes
place among two chiral Schwinger models\cite{JR} having opposite chiralities to
reproduce the usual Schwinger model\cite{J}. Chiral models satisfy unitarity provided $a_\pm\geq 1$ and
the spectrum consists of a vector boson with mass, $m^2 = \frac{e^2 a^2}{a-1}$ 
and a massless chiral boson.  The values of the parameters obtained here
just saturate the bound: the vector boson 
becomes infinite so that it goes out of the spectrum. Thus the
soldering mechanism shows how the massless modes in the chiral Schwinger
models are fused to generate the massive mode of the Schwinger model.
It may be
observed that the soldering process can be carried through for the
non Abelian theory as well, and a relation analogous to (\ref{110s}) is
obtained.

Naively it may appear that the soldering of the left and right chiralities
to obtain a gauge invariant result is a
simple issue since adding the classical Lagrangians
$\bar\psi\Dslash_+\psi$ and $\bar\psi\Dslash_-\psi$, with identical
fermion species, just yields the
usual vector Lagrangian $\bar\psi\Dslash\psi$. The quantum considerations 
are, however, much involved. The chiral determinants, as they occur,
cannot be even defined
since the kernels map from one chirality to the other so that there is no
well defined eigenvalue problem. This is circumvented by
working with
$\bar\psi(i\dslash + e\aslash_{\pm})\psi$, that satisfy an eigenvalue
equation, from which their determinants may be computed. But now a simple
addition of the classical Lagrangians does not reproduce the expected
gauge invariant form. At this juncture, the soldering process becomes
important. It systematically combined the quantized (bosonized)
expressions for the opposite chiral components of {\it different}
fermionic species. The importance of this will 
become more transparent when the three dimensional case is discussed.

Is interesting to examine the impact that different choices
of regularizations may have over our result.  There are two possible situations left.
Here we examine the present regularization and show that different choices for the
parameters $a_\pm$ in (\ref{110s}) lead to the Thirring model.
An alternative regularization prescription leading
to a diverse constraint structure was examined in \cite{AINW}.
Indeed it is precisely
when the mass term exists ($i.e., \,\, a_+ + a_- -2\neq 0$), that (\ref{110s})
represents
the Thirring  model. Consequently, this parametrization complements that used
previously to obtain the vector gauge invariant structure.
The term in parentheses in (\ref{110s}) corresponds to $\bar\psi
(i\dslash +e\aslash) \psi$ so that integrating out the auxiliary $A_\mu$ field
yields,
\be
{\cal L}=\bar\psi i\dslash\psi - \frac{g}{2}(\bar\psi\gamma_\mu\psi)^2\,\,\,\,
;\, g=\frac{4\pi}{a_+ + a_- -2}
\label{A1}
\ee
which is just the Lagrangian for the usual Thirring model. It is known\cite{SC} that 
this model is meaningful provided the coupling parameter satisfies the 
condition $g>-\pi$, so that, $\mid a_+ + a_- \mid >2 $.
This condition complements the condition found earlier.

We have therefore explicitly derived expressions for the chiral determinants
(\ref{30s}) 
which simultaneously preserve the factorization property  and 
gauge invariance of the vector determinant.  It was also perceived that the 
naive way of interpreting the chiral determinants as $W[A_+,0]$ or 
$W[0,A_-]$ led to the supposed incompatibility of factorization with gauge 
invariance. Perturbatively we show the lacking of crossing graphs.  Classically 
these graphs do vanish ($P_+ P_- =0$) so that it becomes evident that this 
incompatibility originates from a lack of properly accounting for the quantum 
effects.  It is possible to interpret this effect, as we will now show,
as a typical quantum mechanical 
interference phenomenon, closely paralleling the analysis in Young's double 
slit experiment.  We also provide a new interpretation for 
the Polyakov-Wiegman\cite{PW}
identity. Rewriting (\ref{140s}) in Fourier space as
\begin{eqnarray}
\label{240s}
W[A_\pm]
= -{N\over 2}\int dk {\mid}\sqrt{k_-\over k_+}A_+ -
\sqrt{k_+\over k_-}A_-
{\mid^2}
\end{eqnarray}
\noindent immediately displays the typical quantum mechanical interference 
phenomenon, in close analogy to the optical example.
The dynamically  
generated mass arises from the interference 
between these movers, thereby preserving gauge invariance. Setting either 
$A_+$ or $A_-$ to vanish, destroys 
the quantum effect, very much like closing one slit in the optical 
experiment destroys the interference pattern.   
Although this analysis was done for the Abelian theory, it is
straightforward to perceive that the effective action for a
non Abelian theory can also be expressed in the form of an absolute
square (\ref{240s}), except that there will be a repetition of
copies depending on the group index.
This happens because only the two-legs graph has
an ultraviolet divergence, leading to the interference (mass) term.
The higher legs graphs are all finite, and satisfy the naive
factorization property.


\subsection{Self-Dual Models and 3D Bosonization.}
While the bosonization in 3D is not
exact, nevertheless, for massive fermionic models in the large mass or,
long wavelength limit, well defined local expressions
are known to exist\cite{C,RB}. These expressions exhibit a self
or an anti self dual symmetry that is dictated by the signature of the fermion
mass. Clearly, this symmetry simulates the dual aspects of
the left/right chiral symmetry in the two dimensional example,
providing a novel testing ground for our ideas. Indeed, two distinct
massive Thirring models with opposite mass signatures,
are soldered to yield a massive Maxwell theory. This result
is vindicated by a direct comparison of the current correlation functions
obtained before and after the soldering process.

To effect the soldering consider the bosonization of the massive Thirring
model in three dimensions\cite{C,RB}. The
relevant partition functional,
in the Minkowski metric is
\be
\label{160}
Z\!\!=\!\!\!\int\!\! D\psi\! D\bar\psi\, e^{i\int d^3x\left
[\bar\psi(i \dslash + m )\psi -\frac{\lambda^2}{2}
j_\mu j^\mu \right]}
\ee
where $j_\mu=\bar\psi\gamma_\mu\psi$ is the fermionic current.
The four fermion interaction is eliminated by introducing an auxiliary
field,
\be
\label{170}
Z\!\!=\!\!\!\int\!\! D\psi\! D\bar\psi\! Df_\mu \, e^{i\int d^3x\left
[\bar\psi\left(i \dslash + m +\lambda \fslash\right)
\psi +\frac{1}{2} f_\mu f^\mu\right]}
\ee
Contrary to the two dimensional models, the fermion integration cannot be
done exactly. Under large mass limit, however, this integration
is possible leading to closed and local expressions\cite{RB1}.
The leading term in this
limit was calculated \cite{DJT} 
and shown to yield the Chern-Simons
three form. Thus the partition functional for the massive Thirring model in
the large mass limit is 
\be 
\label{180}
Z\!\!=\!\!\!\int\!\! Df_\mu \, e^{i\int d^3x\:\left({\lambda^2\over{8\pi}}{m\over{\mid m\mid}}
\epsilon_{\mu\nu\lambda}f^\mu\partial^\nu f^\lambda +
\frac{1}{2} f_\mu f^\mu \right)}
\ee
where the signature of the topological terms is dictated by the corresponding
signature of the fermionic mass term. 
The Lagrangian in the above partition
function defines a self dual model introduced earlier\cite{TPN}. The massive
Thirring model, in the relevant limit, bosonizes to a self dual
model. It is useful to clarify the meaning of this self duality. The 
equation of motion is given by, $f_\mu =-{\lambda^2\over{4\pi}}{m\over{\mid m\mid}}
\epsilon_{\mu\nu\lambda}\partial^\nu f^\lambda$
from which the following the relations $\partial_\mu f^\mu =0$ and
$\left(\partial_\mu\partial^\mu + M^2\right)f_\nu = 0;\,\, M=\frac{4\pi}{\lambda^2}$
are verified.
A field dual to $f_\mu$ is defined as, $\mbox{}^* f_\mu = {1\over M} \epsilon_{\mu\nu\lambda}\partial^\nu f^\lambda$
where the mass parameter $M$ is inserted for dimensional reasons. Repeating
the dual operation, we find, $\mbox{}^*{\left(\mbox{}^*{f_\mu}\right)}= 
{1\over M} \epsilon_{\mu\nu\lambda}\partial^\nu\,\,\mbox{}^*{f^\lambda}=f_\mu$
thereby validating the definition of
the dual field.  Combining these results
we conclude that, $\mbox{}^*f_\mu=- \frac{m}{\mid m \mid} f_\mu$.
Hence, depending on the sign of the fermion mass term, the bosonic theory
corresponds to a self or an anti self-dual model.  Likewise, the Thirring
current bosonizes to the topological current
$j_\mu = \frac{\lambda}{4\pi}\frac{m}{\mid m\mid}\epsilon_{\mu\nu\rho} \partial^\nu f^\rho$.

The close connection with the two dimensional analysis is now evident.
There the starting point was to consider two distinct fermionic theories with 
opposite chiralities. In the present instance, the analogous thing is to
take two independent Thirring models with identical coupling strengths but
opposite mass signatures,
\ba
\label{240}
{\cal L_\pm}&=&\bar\psi_\pm\left(i\dslash \pm m_\pm\right)\psi_\pm -\frac{\lambda^2}{2}\left(\bar\psi_\pm\gamma_\mu\psi_\pm\right)^2
\ea
Note that only the relative sign between the mass parameters is important,
but their magnitudes are different. From now on it is also assumed that
both $m_\pm$ are positive.
Then the bosonized Lagrangians are, respectively,
\ba
\label{250}
{\cal L_\pm}&=&\frac{\pm 1}{2M} 
\epsilon_{\mu\nu\lambda}f_\pm^\mu\partial^\nu f_\pm^\lambda +
{1\over 2} \eta_{\mu\nu}f_\pm^\mu f_\pm^\nu
\ea
where $f_\pm^\mu$ are the distinct bosonic vector fields.

It is now possible to effect the soldering following the general prescription detailed in the last section.  The final result, after elimination of auxiliary soldering fields, is
\ba
\label{350}
{\cal L}_S =  - \frac{1}{4} F_{\mu\nu}F^{\mu\nu} + 
{M^2\over 2}A_\mu A^\mu
\ea
where, $A_\mu = {1\over{\sqrt{2} M}}\left(f^+_\mu - f^-_\mu\right)$
is the usual field tensor expressed in terms of the basic entity $A_\mu$.
The soldering mechanism has precisely fused
the self and anti self dual symmetries to yield a massive Maxwell.

We conclude, therefore, that two massive Thirring models with opposite 
mass signatures, in the long wavelength limit,
combine by the process of bosonization and soldering, to a massive
Maxwell theory. The bosonization of the composite current, obtained
by adding the separate contributions from the two models, is given in
terms of a topological current of the massive vector theory.
These results cannot be obtained by a
straightforward application of conventional bosonization techniques.
The massive modes in the original Thirring models are 
manifested in the two modes of (\ref{350}) so that there is a proper
matching in the degrees of freedom.

\section{ Electromagnetic Duality in Different Dimensions.}

\subsection{Duality Groups}
The problem with the space-time dimensionality is a crucial one.
The distinction among the different even dimensions is manifest by the
following double duality relation,
\begin{equation}
\label{dg10}
\mbox{}^{**}F = \begin{cases}
+F,& if $\;\;D=4k+2$\cr
	-F,& if $\;\;D=4k$
\end{cases}
\end{equation}
where $*$ denotes the usual Hodge operation and $F$ is a $\frac D2$-form.  The concept of self duality seems to be well defined only in twice odd dimensions, and not present in the twice even cases and (\ref{dg10}) apparently leads to
separate consequences regarding the duality groups in these cases. 
The invariance of the actions in different
$D$-dimensions is preserved by the following groups,
\begin{equation}
\label{dg20}
{\cal G}_d= \begin{cases}
Z_2,& if $\;\;D=4k+2$\cr
	SO(2),& if $\;\;D=4k$
\end{cases}
\end{equation}
which are called the ``duality groups".
The duality operation is characterized by a one-parameter SO(2) group of symmetry in D=4k dimensions, while for D=4k+2 dimensions it is manifest by a discrete $Z_2$ operation. Notice that only the 4 dimensional Maxwell theory and its 4k extensions would possess duality as a symmetry, while for the 2 dimensional scalar theory and its 4k+2 extensions duality is not even definable. 
We shall discuss the physical origin of this dichotomy.

A solution for the problem came with the recognition of a 2-dimensional internal structure hidden in the space of potentials\cite{DZ,DT,SS}.  
Recently this author\cite{CW23} has developed a systematic method for obtaining and investigating different aspects of duality symmetric actions that embraces all dimensions. Later, with collaborators, these methods were extended to massive fields of arbitrary ranks\cite{p-forms}.
A redefinition of the fields in the first-order form of the action naturally discloses the 2-dimensional internal structure hidden into the theory.  This procedure produces two distinct classes of dual theories characterized by the opposite signatures of the (2x2) matrices in the internal space.  These actions correspond to self dual and anti-selfdual representations of the original theory.
Indeed the dichotomy (\ref{dg20}) seems to be of much deeper physical origin, since it attributes
different group structures to distinct dimensions and ranks. 

The dual projection operation, that systematically discloses the internal duality space of any theory in D-dimensions is quickly discussed for massless tensors.
The distinction of the duality groups is manifest, in the dual projection approach by the following construction.  The first-order action for a free field theory is, in general, given as
\be
\label{dg70}
{\cal L} = \Pi \cdot \dot\Phi - \frac 12 \Pi\cdot\Pi - \frac 12 
\partial\Phi\cdot\partial\Pi
\ee
with $\Pi$ and $\Phi$ being generic free tensor fields in D-1 dimensions and $\partial$ an appropriate differential operator.
For visual simplicity we omit all tensor and space-time indices describing the fields, unless a specific example is considered.
The parity of $\partial$ has a particularly interesting dependence with the dimensionality,
\begin{equation}
\label{dg80}
P\left(\partial\right) = \begin{cases}
+ 1,& if $D=4k$\cr
	- 1,& if $D=4k+2$
\end{cases}
\end{equation}
where parity is defined as,
\be
\label{dg90}
\int \Phi \cdot\partial\Psi = P\left(\partial\right)\int \partial\Phi\cdot\Psi
\ee
Take for instance the specific cases of 2 and 4 dimensions where $\partial$ is defined as,
\begin{equation}
\label{dg100}
\partial = \begin{cases}
\partial_x,& if $\;\;D=2$\cr
	\epsilon_{kmn}\partial_n,& if $\;\;D=4$
\end{cases}
\end{equation}
which we recognize as odd and even parity respectively.
The internal space is disclosed by a suitable field redefinition that is dimensionally dependent,
\begin{eqnarray}
\label{dg110}
\Phi & \rightarrow & A_+ + A_-\nonumber\\
\Pi &\rightarrow & \eta\left(\partial A_+ - \partial A_-\right)
\end{eqnarray}
with $\eta=\pm$ defining the signature of the canonical transformation.  The effect of the dual projection (\ref{dg110}) into the first order action is manifest as,
\begin{equation}
\label{dg120}
{\cal L} = \left(\dot A_\alpha\sigma_3^{\alpha\beta}\partial A_\beta
+ \dot A_\alpha\epsilon^{\alpha\beta}\partial A_\beta\right) - \partial A_\alpha\cdot\partial A_\alpha
\end{equation}
We can appreciate the impact of the dimensionality over the structure of the internal space, and the
role of the operator's parity in determining the appropriate group for each dimension.  For twice odd dimension the dual projection produces the diagonalization of the first-order action into chiral actions, while for twice even dimensions, the result is either a self or an anti-selfdual action, depending on the sign of $\eta$,
\begin{equation}
\label{dg130}
{\cal L} = \begin{cases}
\eta\dot A_\alpha\sigma_3^{\alpha\beta}\partial A_\beta - \partial A_\alpha\cdot\partial A_\alpha
  ,& if $\;\;D=4k+2$\cr
	\eta\dot A_\alpha\epsilon^{\alpha\beta}\partial A_\beta - \partial A_\alpha\cdot\partial A_\alpha
,& if $\;\;D=4k$
\end{cases}
\end{equation}
By inspection on the above actions one finds that while the first is duality symmetric under the discrete $Z_2$, the second is invariant under the continuous one-parameter group $SO(2)$.  This is in accord with general discussion based on algebraic methods\cite{DGHT}. In fact, using the choice of operators in (\ref{dg100}) we easily find that the D=2 case describes a right and a left Floreanini-Jackiw
chiral actions\cite{FJ} if we identify $\Phi$ with a scalar field,
\begin{equation}
\label{dg136}
{\cal L}= \eta\dot A_\alpha\sigma_3^{\alpha\beta} A'_\beta -  A'_\alpha\cdot A'_\alpha
\end{equation}
The second case, on the other hand, describes either selfdual or anti-selfdual Schwarz-Sen actions, according to the signature of $\eta$,
\begin{equation}
\label{dg138}
{\cal L}=\eta\dot \varphi_k^\alpha\epsilon^{\alpha\beta}B_k^\beta - B_k^\alpha\cdot B_k^\alpha
\end{equation}
if we identify $\Phi$ with a vector field, $\Phi \rightarrow \varphi_k^\alpha$ and $\partial\Phi\rightarrow \epsilon_{kmn}\partial_m \varphi_n^\alpha = B_k^\alpha$, with $B_k^\alpha$ being the magnetic field.

In summary the above analysis clearly shows that the physical origin for the dimensional dependence of the electromagnetic duality group lies in the parity dependence on dimensionality of a curl-operator naturally defined in the solution of the Gauss law.  For the massive case \cite{p-forms} the analysis follows analogously but the physical origin of the curl-operator is quite distinct. These features are summarized in the following table, 
\begin{center}
\begin{tabular}{|c|c|c|c|c|}
\hline
dimension  & massive &  massless  &  rank  & dof    \\ \hline
13 &  SO(2)   &         &  6  & 924   \\ \hline
12 &          &  SO(2)  &  5  &  252    \\ \hline
11 &  $Z_{2}$   &         &  5  &  252    \\ \hline
10 &          &  $Z_{2}$  &  4  &   70    \\ \hline
9  &  SO(2)   &         &  4  &   70    \\ \hline
8  &          &  SO(2)  &  3  &   20    \\ \hline
7  &  $Z_{2}$   &         &  3  &   20    \\ \hline
6  &          &  $Z_{2}$  &  2  &   6     \\ \hline
5  &  SO(2)   &         &  2  &   6     \\ \hline
4  &          &  SO(2)  &  1  &   2     \\ \hline
3  &  $Z_{2}$   &         &  1  &   2     \\ \hline
2  &          &  $Z_{2}$  &  0  &   1     \\ \hline
\end{tabular}
\end{center}

\subsection{Duality Equivalence: Noether Embedding.}
Recently we proposed a new technique to perform duality mappings for vectorial models in any dimensions that is alternative to the master action approach\cite{IW11}.
This technique is based on a two-fold approach that simultaneously lift a global symmetry in its local form and may be realized by an iterative embedding of Euler vectors counter-terms.  The use of Euler vectors is done to guarantee the dynamical equivalence between the models, while the embedding algorithm progressively subtracts the gauge offending terms from the theory\cite{constraints}.
A variation of this technique is exploited here since it seems to be the most appropriate for non-Abelian generalizations of the dual mapping concept.   It seems however that for the nonabelian case the lifting to a local symmetry conflicts with dual equivalence and either one must be sacrificed. In this work we will look for duality equivalence in detriment of gauge symmetry.  In another contribution we followed the opposite route and construct a gauge invariant theory out of the gauge variant model \cite{IW11}
 
Using the well known equivalence between the self-dual\cite{TPN} and the topologically massive models\cite{DJT} proved by Deser and Jackiw\cite{DJ} through the master action approach, a correspondence has been established between the partition functions for the MTM and the Maxwell-Chern-Simons (MCS) theories.  The situation for the case of fermions carrying non-Abelian charges, however, is less understood due to a lack of equivalence between these vectorial models, which has only been established for the weak coupling regime\cite{BFMS}.  As critically observed in \cite{KLRvN}, the use of gauge invariant master actions in this situation is ineffective for establishing dual equivalences. We shall establish an exact result showing that these two nonabelian models {\it are not} equivalent\cite{notdual}.  We will then find explicitly the dual equivalent for the nonabelian self-dual model and discuss its differences and similarities with the YMCS model.

 First, let us discuss the general idea in \cite{IW11,AINRW} and consider the duality mapping from a general gauge variant model described by a Lagrangian density ${\cal L}_0$ and call its dual equivalent as $\mbox{}^*{\cal L}$.
 As shown in \cite{AINRW}, the final effect of the gauge embedding algorithm is materialized in the following form
 \be
 \label{p10}
 {\cal L}_0[A] \to \mbox{}^*{\cal L}[A] = {\cal L}_0[A] + f(K^2)
 \ee
 where the Euler kernel 
 \be
 \label{p20}
 \delta{\cal L}_0[A] = K_\mu[A]\, \delta A^\mu
 \ee
 defines the classical dynamics of the original theory.  Therefore equivalence will be impacted if the final function is demanded to satisfy the condition
 \be
 \label{p30}
 f(K^2)\mid_{K=0} = 0
 \ee
 The embedding technique was originally explored in the context of the
 soldering formalism \cite{ABW,BW} and explored to study equivalent dynamics of vector models in diverse regimes that includes non-linearities and couplings to dynamical matter. The variation described above seems to be an appropriate technique for non-nonabelian generalization of the dual mapping concept. In general the algorithm works by demanding gauge invariance but we will relax this condition here and choose a simple function in (\ref{p10}) leading to duality just by inspection.

Using the gauge embedding idea, we clearly find the dual equivalence for the non-Abelian self-dual model, extending the program proposed by Deser and Jackiw in the Abelian domain.
These results have consequences for the bosonization identities from the massive Thirring model
into some ``topologically massive model", which are considered here, and also allows for the extension of the fusion of
the self-dual massive modes\cite{BW} for the non-Abelian case\cite{IW11}.

The non-Abelian version of the
vector self-dual model (\ref{180}), which is our main concern, is given by
 \ba
 \label{pp220}
 {\cal S}_{SD}[A] = -\frac N{g^2}\int d^3 x\: tr \left(A_\mu A^\mu \right) - \frac \chi{8\pi} {\cal S}_{CS}[A]&&\nonumber\\
 \!\!\!\!= \!\!\int\!\! d^3 x\!\!\left[\! \frac N{2g^2}  A^a_\mu A^{a\mu}\!\! +\!\! \frac \chi{8\pi} \epsilon^{\mu\nu\lambda}\!\!\!
 \left(\!\!A_\mu^a \partial_\nu A_\lambda^a\!\!+\!\!\frac {f^{abc}}{3}A_\mu^a A_\nu^b A_\lambda^c \!\!\right)\!\!\right]&&\nonumber
 \ea
where ${A}_{\mu} = A_{\mu}^{a}{t}^{a}$, is a vector field taking values in the Lie algebra of a symmetry group $G$ and ${ t}^{a}$ are the matrices
representing the underlying non Abelian gauge group with $a= 1,\ldots , \mbox{dim}\:G$.

Using the master action approach, the NASD model has been shown to be equivalent to the gauge invariant Yang-Mills-Chern-Simons (YMCS) theory in the weak coupling limit $g\to 0$ so that the Yang-Mills term effectively vanishes.
Here we are using the bosonization nomenclature that relates the Thirring model coupling constant $g^2$ with the inverse mass of the vector model.
To study the nonabelian dual equivalence for all coupling regimes, a problem open for more than twenty years, and the consequences over the bosonization program is a contribution of this program.

 We are now in position to prove our main result -- the actual computation of the dual equivalent to the nonabelian self-dual model.  To do this we need to compute the Euler vector to the model
 \be\label{PP230}
 K_\mu^a = \frac N{g^2}  A^a_\mu  + \frac \chi{8\pi} \epsilon^{\mu\nu\lambda}
 F_{\nu\lambda}^a
 \ee
 and choose a function $f(K^2)$ so that the gauge offending term gets subtracted out.  It is simple to see that the following choice will do,
 \ba
 \label{pp240}
 &&\mbox{}^*{\cal S}[A] \!=\! {\cal S}_{SD}[A] - \frac {g^2}{2 N} \int d^3 x\: K_\mu^a K^{\mu a}\nonumber\\
 &=& \!\!\!\frac 1{8\pi}\!\! \int\!\! d^3 x\! \left\{\! -\frac {g^2}{8\pi N} F_{\mu\nu}^a F^{\mu\nu a}\right.\nonumber\\
 &-& \left.\chi \epsilon^{\mu\nu\lambda}\!\!\left(\!A_\mu^a \partial_\nu A_\lambda^a\!\!+\!\!\frac 23{f^{abc}}A_\mu^a A_\nu^b A_\lambda^c\!\right) \! \right\}
 \ea
 leading to a Yang-Mills type theory.  Notice that the second term is Chern-Simons like term but not quite the CS itself.  By construction both theories have the same dynamics but the dual model is still not gauge invariant. Although we start by subtracting the gauge variant piece, the counter-term added has broken the CS term spoiling its gauge invariance for small gauge transformations. To produce a gauge invariant model, at least for those transformations connected to the identity, one has to add a counter-term not proportional to the Euler kernel which will eventually spoil dynamical equivalence.  It becomes then quite clear that the YMCS model is not dual to the nonabelian self-dual model.  The model in (\ref{pp240}) is and this is an exact result.

In summary, a general method has been developed that establishes dual equivalence between self-dual and topologically massive theories based on the idea of gauge embedding over second-class constrained systems.  The equivalence has been established using an adaptation of the iterative Noether procedure both for Abelian and non-Abelian self-dual models, including the cases with coupling to dynamical charged fermions.
In the process we have also shown the duality transformation that correctly defines the inherent self-duality property
of the noninvariant actions.

\end{document}